\documentclass[a4paper, times, 10pt,twocolumn]{article}
\usepackage[top=4.9cm,bottom=3.7cm,left=1.5cm,right=1.5cm]{geometry}
\usepackage{ICMLC}
\usepackage{times}
\usepackage{graphicx}
\usepackage{indentfirst}
\usepackage{latexsym}
\usepackage{amsmath}
\usepackage{algorithm}
\usepackage{algorithmic}
\usepackage[margin=8pt,font=footnotesize,labelfont=bf,labelsep=period
]{caption}

\pdfpagewidth=\paperwidth
\pdfpageheight=\paperheight
\begin{document}

\title{LEARNING COMMUNICATION BETWEEN HETEROGENEOUS AGENTS IN MULTI-AGENT REINFORCEMENT LEARNING FOR AUTONOMOUS CYBER DEFENCE}  

\author{\bf{\normalsize{ALEX POPA${^1}$, ADRIAN TAYLOR${^2}$, RANWA AL MALLAH${^3}$}}\\ 
\\
\normalsize{$^1$Department of Electrical and Computer Engineering, Royal Military College of Canada, Kingston Ontario, Canada}\\
\normalsize{$^2$Defence Research and Development Canada, Ottawa Ontario, Canada}\\
\normalsize{$^3$Department of Computer Engineering and Software Engineering, Polytechnique Montréal, Montréal Québec, Canada} \\
\normalsize{E-MAIL: alex.popa@rmc-cmr.ca, adrian.taylor@drdc-rddc.gc.ca, ranwa.al-mallah@polymtl.ca}\\
\\}

\maketitle \thispagestyle{empty}

\begin{abstract}
   {Reinforcement learning techniques are being explored as solutions to the threat of cyber attacks on enterprise networks. Recent research in the field of AI in cyber security has investigated the ability of homogeneous multi-agent reinforcement learning agents, capable of inter-agent communication, to respond to cyber attacks. This paper advances the study of learned communication in multi-agent systems by examining heterogeneous agent capabilities within a simulated network environment. To this end, we leverage CommFormer, a publicly available state-of-the-art communication algorithm, to train and evaluate agents within the Cyber Operations Research Gym (CybORG). Our results show that CommFormer agents with heterogeneous capabilities can outperform other algorithms deployed in the CybORG environment, by converging to an optimal policy up to four times faster while improving standard error by up 38\%. The agents implemented in this project provide an additional avenue for exploration in the field of AI for cyber security, enabling further research involving realistic networks.}
\end{abstract}

\begin{keywords}
   {MARL; CybORG; CommFormer; AI for security; Cyber security; Inter-agent communication}
\end{keywords}

\Section{Introduction}



Modern computer and digital systems exist in a world of constant inter-connectivity where cyber threats pose a persistent risk to the daily lives of users \cite{national_cyber_threat_assessment_2025-2026}. The situation is further complicated by the rapid evolution of tactics, techniques and procedures employed by malicious actors operating in the cyber domain. State-affiliated actors from great and middle powers across the world are increasingly leveraging AI and advanced techniques to circumvent traditional defences \cite{crowdstrike_2025}. Recent reporting from 2024 indicates sustained targeting of critical infrastructure, highlighting a shift toward more sophisticated, large-scale campaigns against rival national interests \cite{national_cyber_threat_assessment_2025-2026}\cite{crowdstrike_2025}.

To counter these evolving threats, researchers have looked toward Machine Learning (ML) as a means to secure the cyber domain. While Reinforcement Learning (RL) has proven effective for Network Intrusion Detection Systems (NIDS) in identifying malware and Distributed Denial of Service (DDoS) attacks, single-agent RL architectures have had limited success in responding to large-scale complex attacks \cite{contractor}\cite{nguyen}. Multi-agent Reinforcement Learning (MARL) offers a scalable alternative by distributing detection and response across the network. MARL agents can be trained to replace or support an existing Host-based Intrusion Detection System (HIDS), greatly extending the AI for security use case. While MARL algorithms provide a myriad of benefits \cite{lanctot}, they often suffer from environmental instability \cite{hu}.

Inter-agent communication has been proposed as a means to mitigate MARL instability \cite{hu}. Building upon the foundations established in \cite{contractor}, this work explores the application of MARL agents with heterogeneous capabilities, meaning varied observation and action spaces. By employing CommFormer, a state-of-the-art communication algorithm \cite{hu}, this work demonstrates that heterogeneous communication agents can outperform baselines established in the field \cite{contractor}.

\Section{Related Work}

MARL has been described as a natural extension of single-agent RL, where agents engaging in a shared environment learn to map states to actions through an iterative trial-and-error process \cite{contractor}\cite{hu}. MARL applications have been aimed at addressing scalability limitations associated with single-agent architectures. This increase in agent numbers generally leads to dynamic uncertainty and environment instability \cite{hu}. To mitigate this, researchers have used approaches like Centralized Training Decentralized Execution (CTDE), which allows agents to share global information during training while operating independently during execution. Inter-agent communication further enhances coordination, particularly in partially observable environments where agents have limited observation spaces and must learn to share vital data. Despite the advantages associated with communication, drawbacks exist, such as increased bandwidth usage, slowdowns at inference time, and additional complexity in algorithm architecture.

In the context of AI for cyber security, MARL agents contend with partial observability, sparse rewards, and high stochasticity. For instance, a host-based agent is limited to local observations, requiring communication to coordinate or support a network-wide response \cite{sukhbaatar}. Additionally, Autonomous Cyber Defence (ACD) requires multi-step responses to malicious actions, leading to delayed feedback for correct choices. Finally, the stochasticity of computer networks, where network traffic and benign user actions introduce significant noise, leads to a difficult learning environment. Despite these challenges, MARL agents can be tailored to niche roles, helping minimize the need for computational power, as well as to leverage advanced  communication algorithms, capable of limiting bandwidth requirements.

The Reinforced Inter-Agent Learning (RIAL) algorithm, utilizing independent Q-networks, and the Differentiable Inter-Agent Learning (DIAL) \cite{DIAL} algorithm, enabling gradient sharing across agents, introduced deep Q-networks for communication. Complementary Attention for Multi-Agent reinforcement learning (CAMA) \cite{CAMA}, also using Q-Learning, attempts to optimize agent observation spaces by dynamically altering agent observation spaces, a feature not currently supported by CybORG.

Communication Neural Network (CommNet) \cite{CommNet}, using the REINFORCE training algorithm \cite{REINFORCE}, builds a channel where agents transmit continuous vectors. CommNet's multiple communication cycles per timestep demand excessive bandwidth, making it ill-suited for applications in ACD. Individualized Controlled Continuous Communication Model (IC3Net) \cite{IC3Net}, also trained via the REINFORCE algorithm, improves efficiency by introducing a gating mechanism which allows agents to learn when to communicate. While useful for mixed cooperative-competitive settings, IC3Net lacks demonstrated support for heterogeneous agents.

Bidirectionally-Coordinated Network (BiCNet) \cite{bicnet}, an early actor-critic example, requires the global state to be provided at each timestep, an unrealistic expectation in partially observable environments. Targeted Multi-Agent Communication (TarMAC) \cite{tarmac} utilizes an actor-critic architecture with a signature-based soft-attention mechanism, enabling agents to weigh message importance and target recipients. Similarly, Distributed Targeted Multi-Agent Communication (DTMAC) \cite{dtmac} adds historical message support. While highly effective, both algorithms were tested only on homogeneous agents.

Multi-Agent Incentive Communication (MAIC) \cite{maic} employs teammate modelling and Q-Learning to generate incentive messages that directly bias teammates' local Q-value tables. Although MAIC supports heterogeneous agents, this abstract communication lacks interpretability.

Advanced graph-based approaches include Multi-Agent Graph-attentIon Communication (MAGIC) \cite{magic}, Information Maximizing Gated Spares Multi-Agent Communication (IMGS-MAC) \cite{imgs-mac} and CommFormer \cite{hu}. MAGIC employs Graph Attention Networks (GATs) via a Scheduler and Message Processor to manage multi-round communication. IMGS-MAC enforces sparsity to maximize information sharing within strict bandwidth constraints. Despite their architectural advantages, these graph models lack publicly available code and verified support for heterogeneous agents, ultimately leading to the use of CommFormer.



\Section{Methodology}
We selected the CommFormer algorithm \cite{hu} for our use case as it was found to provide "performance levels comparable to approaches that permit unrestricted information sharing among agents" \cite{hu} while modelling agents in an intuitive way. CommFormer models multi-agent interactions as a Markov game defined by the tuple $\langle \mathcal{N},\boldsymbol{\mathcal{O}},\boldsymbol{\mathbf{A}},\mathcal{R},\mathcal{P},\gamma \rangle$ where $\mathcal{N}$ is the number of agents, $\boldsymbol{\mathcal{O}}$ is the joint observation space, $\boldsymbol{\mathcal{A}}$ is the joint action space, $\boldsymbol{\mathbf{A}}$ maps $\boldsymbol{\mathcal{O}}$ and $\boldsymbol{\mathcal{A}}$ to the reward range, $mathcal{P}$ defines the probability function, and $\gamma$ denotes the discount factor. Inter-agent communication is framed as a learnable directed graph $\mathcal{G}=\langle\mathcal{V},\mathcal{E}\rangle$, optimizing both the graph structure and architectural parameters concurrently via a bi-level optimization process. In this graph, nodes represent agents ($\mathcal{N}$) and edges ($\mathcal{E}$) denote unidirectional communication channels. A sparsity parameter, $\mathcal{S}$, strictly governs the number of active edges $\mathcal{S} \times N^2$, ensuring bandwidth efficiency.

To train its agents, CommFormer employs an encoder-decoder transformer architecture which leverages a self-attention mechanism to calculate scores via Query, Key, and Value vectors, helping capture complex dependencies between agents and observations. Crucially, the learned communication graph masks information from other agents, restricting information flow strictly to authorized channels. Finally, the decoder generates joint action sequences auto-regressively, feeding each generated action back into the decoder to inform the actions of other agents operating in the environment \cite{hu}.

\begin{figure*}[t]
\centering
\includegraphics[width=1\textwidth]{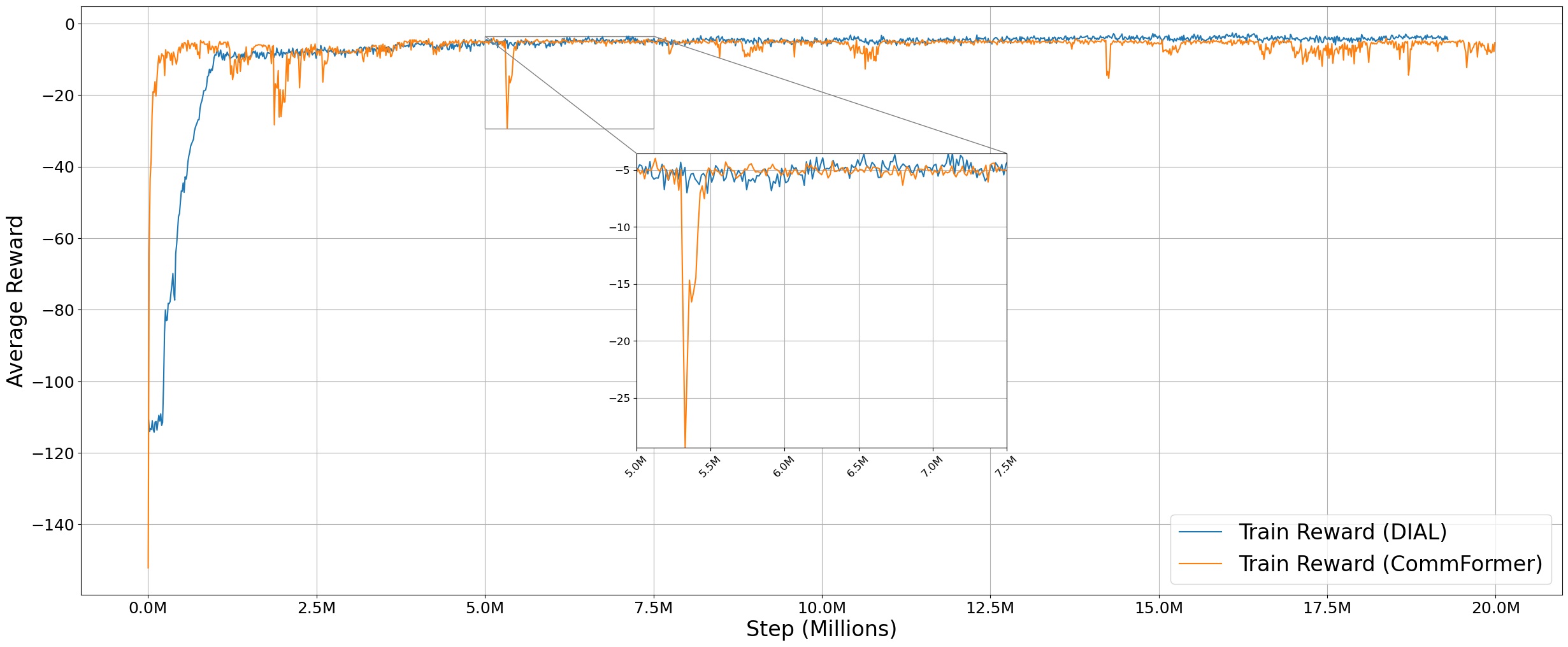}
\caption{Homogeneous agent scenario training curve. The inset plot shows best case scenario performance of approximately -4 average reward.} 
\label{fig:homogeneous_training_curve}
\end{figure*}

\begin{figure*}[t]
\centering
\includegraphics[width=1\textwidth]{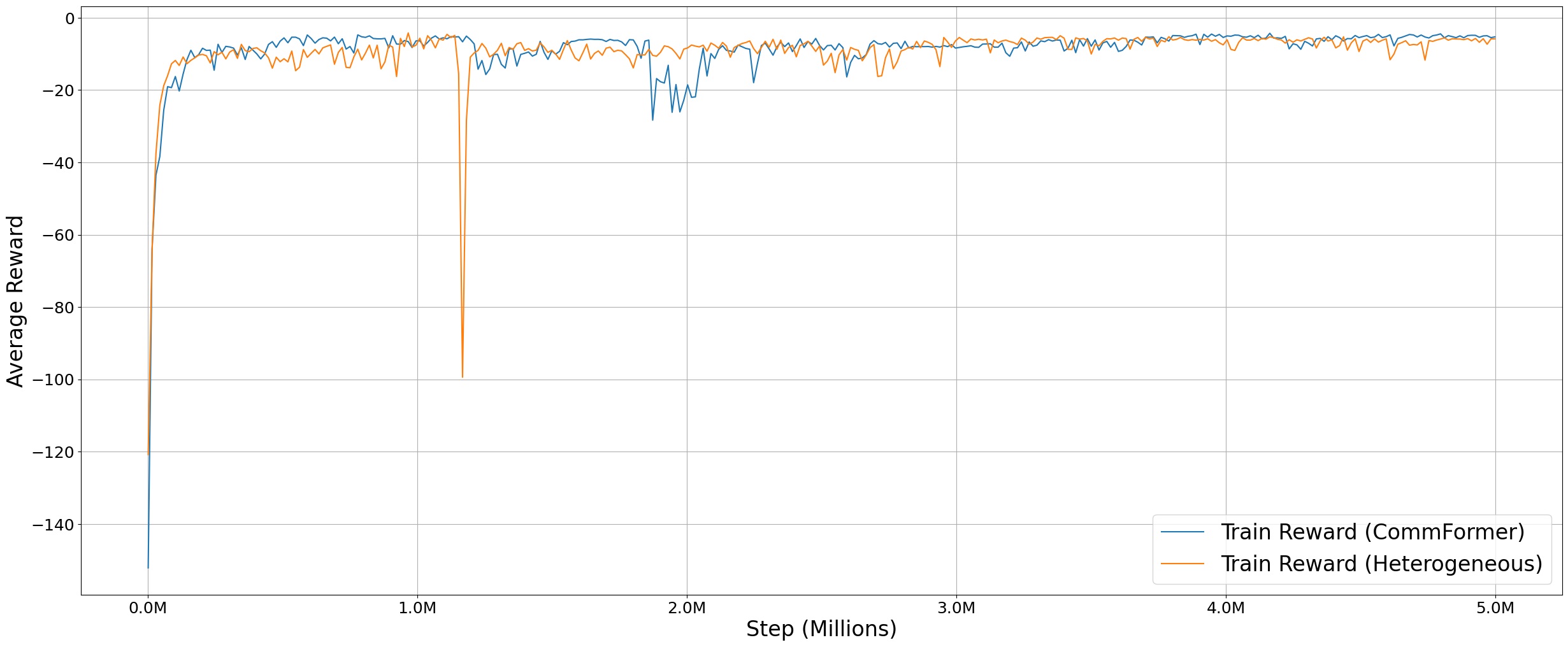}
\caption{Heterogeneous agent scenario training curve.} 
\label{fig:heterogeneous_training_curve}
\end{figure*}


This work utilizes CyMARL, an extension of the CybORG \cite{standen} environment, built upon PyMARL2 \cite{pymarl2}, to train MARL algorithms for ``tactical decision-making in cyber defence'' \cite{wiebe}. Our baseline scenario, features a six-host, two-subnet network where a rigidly programmed red (attacker) agent methodically seeks to exploit vulnerabilities and escalate privileges with the end goal of disrupting a critical server (OpServer).

Defending the network are two blue (defender) agents, acting as subnet Intrusion Prevention Systems (IPS), which must learn to mitigate red incursions through actions like Monitor, Remove, Analyse, Restore, and Block. The training process is guided by a team-based reward function that heavily penalizes the blue agents for allowing compromises to network assets as well as for unnecessarily disruptive defensive actions. To simulate realistic, partially observable conditions, the blue agents receive localized, binary-formatted observations of host and subnet statuses, while automatic monitoring actions only detect exploits with a 50\% success rate, enabling red agents to establish covert, privileged footholds on network hosts. Blue agents are forced to learn indicators of compromise associated with undetected exploits as a result.

CommFormer is implemented using the Python language, its code features a highly modular structure that can be easily adjusted to interact with any environment capable of processing observation, reward and action sequences. The integration of the CommFormer algorithm into CyMARL required the creation of two wrappers. The wrappers developed for this work use the CommFormer Predator-Prey/Predator-Capture-Prey environment scripts created by Hu et al. \cite{hu} as a base to build from. The higher level wrapper is focused on parsing command-line variables, such as hyperparemeters and system information (i.e. CUDA GPU availability and multi-processing settings), helping set up training the environments and saving training data as needed. The underlying wrapper we developped, focuses on the interacting with CyMARL. Its pseudocode can be seen in Algorithm \ref{alg:CybORG_runner} below.

\begin{algorithm}
    \caption{CybORG Runner} \label{alg:CybORG_runner}
    \begin{algorithmic}[1]
        \STATE \textbf{Input:} Arguments and CommFormer configuration data supplied by Train CyMARL CommFormer.
        \STATE \textbf{Initialize:} CybORG runner class using the CommFormer runner class parent.
        \STATE Calculate number of episodes using environment steps and number of multi-processing threads.
        \FOR{episode in episodes}
            \STATE Reset environment and Replay Buffer $\mathcal{B}$.
            \STATE Log communication channel matrix for the episode.
            \FOR{step in training episode length}
                \STATE Sample actions from CommFormer.
                \STATE Observe reward, next observation and available actions from CyMARL.
                \STATE Insert data into buffer $\mathcal{B}$.
            \ENDFOR
            \STATE Compute return and update the CommFormer network.
            \STATE Log information, actions and save model according to pre-defined logging interval.
            \STATE Run evaluation episodes according to pre-defined evaluation interval.
        \ENDFOR
    \end{algorithmic}
\end{algorithm}

Additional changes to both the CyMARL and CommFormer codebases include the addition of support for agents with different observation and action spaces, the ability to log agent actions during training and evaluation, the creation of CybORG environment interaction workers used in multi-processing, as well as numerous tweaks to variables exchanged between CyMARL and CommFormer. The CyMARL-CommFormer repository, including details regarding code changes, is available publicly on GitHub \cite{cymarl_commformer_github}.

\Section{Evaluation}
Following the integration of CommFormer into CyMARL, the algorithm was evaluated across three progressively complex scenarios to validate its performance and scalability against established baselines.

\textbf{Homogeneous Baseline}. To ensure a fair comparison against prior DIAL implementations in CyMARL \cite{contractor}, this initial scenario maintained an identical network topology (six hosts across two subnets), red agent attack path, and uniform blue agent action/observation spaces. Previously optimized hyperparameters outlined in \cite{contractor} and \cite{hu} were used to establish our baseline.

\textbf{Heterogeneous Agent Scenario}. The scenario was then adjusted by introducing a specialized firewall agent (BlueFW). This setup established heterogeneous agents with distinct action and observation spaces. The subnet agents (Blue0 and Blue1) were tasked with monitoring hosts and executing targeted mitigating actions, such as Remove, Restore, and Analyse, while the BlueFW agent was strictly limited to executing the Block action. Zero-padding was introduced to normalize the observation vector sizes across these varied agents.

\textbf{Host-Based Scenario}. To test CommFormer's scalability, the network was expanded to include individual blue agents for each host, simulating localized Endpoint Detection and Response (EDR) applications operating alongside the central BlueFW agent. Because this restricted each agent's observation space to a single host, training instability emerged. Consequently, the reward function was adjusted to increase penalties for unnecessary Analyse actions while discounting correctly coordinated Block actions. Additional hyperparameter tuning (e.g., entropy coefficient, PPO clip) was conducted to prevent convergence on sub-optimal policies.

\begin{figure*}[t]
\centering
\includegraphics[width=1\textwidth]{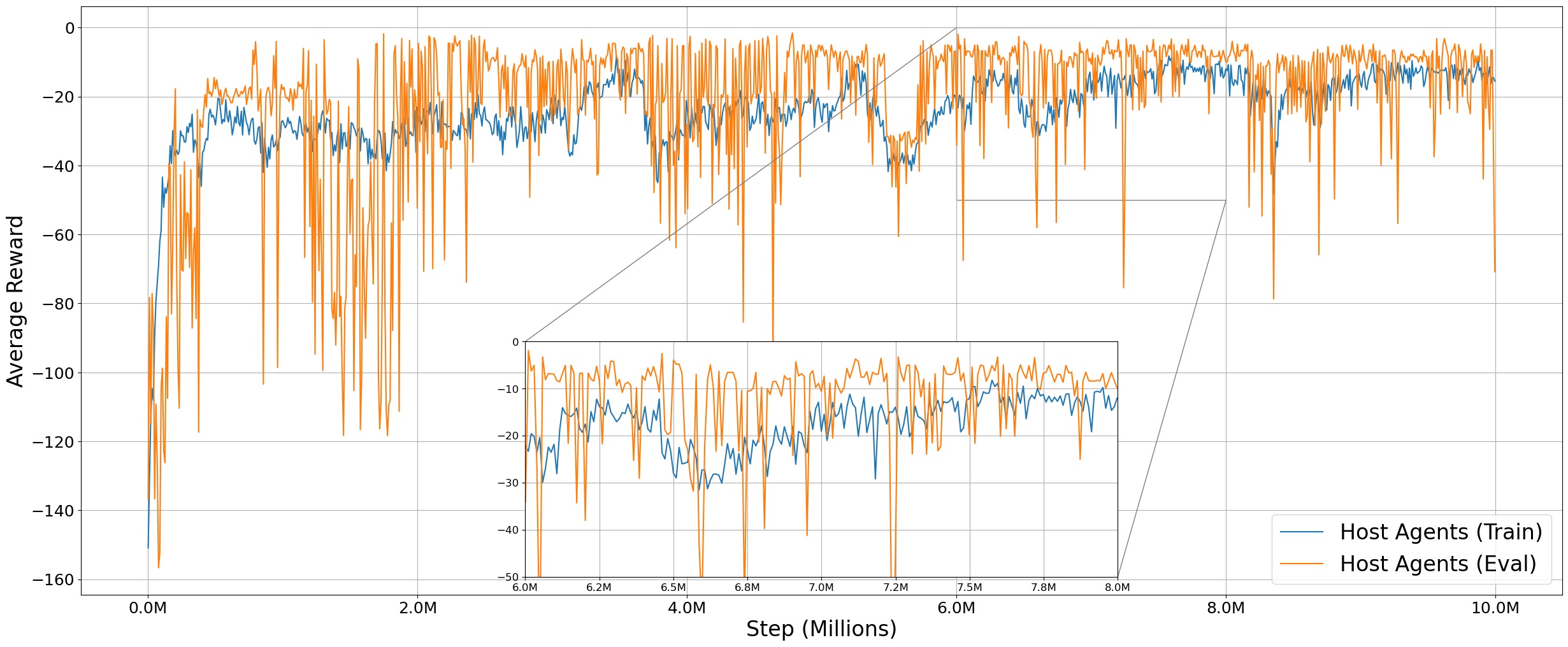}
\caption{Homogeneous agent scenario training curve. The inset plot provides a closer view of peak performance across training and evaluation episodes.} 
\label{fig:host_agents_training_curve}
\end{figure*}

\Section{Results}
\textbf{Homogeneous Baseline}. CommFormer was evaluated throughout a 20 million timestep training session using the homogeneous baseline scenario discussed in the previous section. It demonstrated superior learning efficiency, converging to a semi-optimal policy nearly four times faster than the CyMARL DIAL baseline. CommFormer achieved more stable mean rewards during training $-6.54\pm5.11$ compared to DIAL $-7.9\pm14.39$. Episode logs confirmed this success stemmed from effective communication, as evidenced by an agent detecting a network scan successfully and relaying this data, enabling a peer to execute targeted remediations on a compromised pivot host. Figure \ref{fig:homogeneous_training_curve} illustrates the mean reward assigned to the CommFormer and DIAL agents as they were trained.

\textbf{Heterogeneous Agent Scenario}. CommFormer was subsequently trained and evaluated over 5 million timesteps using the adjusted heterogeneous agent scenario which introduced a specialized firewall agent (BlueFW) alongside subnet agents. The model experienced a minor performance drop (training mean reward of $-9.18\pm8.89$) but maintained rapid convergence. Communication across agents was once again demonstrated when a subnet agent successfully detected a vulnerability scan and alerted the BlueFW agent, whose observation space is limited to the blocked/unblocked status of subnets, prompting immediate network-level mitigation via a Block subnet action. Figure \ref{fig:heterogeneous_training_curve} illustrates the mean reward assigned to the CommFormer heterogeneous agents during training.

\textbf{Host-Based Scenario}. The final scenario simulated localized EDR agents deployed per host, strictly limiting local observations. The severe restrictions placed on CommFormer underscored the necessity for inter-agent communication. While extreme sparsity occasionally delayed initial detections, active channels ultimately routed critical indicators of compromise. Evaluation logs confirmed that isolated host agents successfully initiated Analyse and Restore actions based entirely on observations communicated by peer agents, demonstrating CommFormer's efficacy in decentralized, highly restricted environments. Figure \ref{fig:host_agents_training_curve} illustrates the training and evaluation curves for this scenario.

\Section{Conclusion}

This work successfully adapted the CommFormer algorithm and integrated it into CyMARL to evaluate heterogeneous multi-agent reinforcement learning for ACD. Compared to established DIAL baselines, CommFormer achieved comparable overall performance but converged to near-optimal policies significantly faster. When scaled to complex heterogeneous and highly stochastic host-based scenarios, the agents effectively coordinated mitigating actions across varied observation and action spaces. Although this scalability demonstrated peak performance exceeding baselines, the model exhibited late-stage training instability, necessitating specific hyperparameter and reward tuning. Ultimately, this research validates the efficacy of heterogeneous communicating agents in ACD, establishing a robust foundation for future complex network defence simulations.




\begin{thebibliography}{99}
\bibitem{national_cyber_threat_assessment_2025-2026}
Government of Canada, “National Cyber Threat Assessment 2025-2026”, Canadian Centre for Cyber Security, Oct. 30, 2024. https://www.cyber.gc.ca/en/guidance/national-cyber-threat-assessment-2025-2026
\bibitem{crowdstrike_2025}
“2025 Global Threat Report,” Tech. Rep., CrowdStrike, 2025.
\bibitem{contractor}
F. Contractor and R. Al-Mallah, “Learning to communicate in multi-agent reinforcement learning for autonomous cyber defence,” Master’s thesis, Royal Military College of Canada, 2024.
\bibitem{nguyen}
T. T. Nguyen and V. J. Reddi, “Deep reinforcement learning for cyber security,” IEEE Transactions on Neural Networks and Learning Systems, vol. 34, p. 3779–3795, Aug. 2023.
\bibitem{lanctot}
M. Lanctot, V. Zambaldi, A. Gruslys, A. Lazaridou, K. Tuyls, J. P´erolat, D. Silver, and T. Graepel, “A unified game-theoretic approach to multi-agent reinforcement learning,” Advances in neural information processing systems, vol. 30, 2017.
\bibitem{hu}
S. Hu, L. Shen, Y. Zhang, and D. Tao, “Learning multi-agent communication from graph modeling perspective,” arXiv preprint arXiv:2405.08550, 2024.
\bibitem{DIAL}
J. Foerster, I. A. Assael, N. De Freitas, and S. Whiteson, “Learning to communicate with deep multi-agent reinforcement learning,” Advances in neural information processing systems, vol. 29, 2016.
\bibitem{CAMA}
J. Shao, H. Zhang, Y. Qu, C. Liu, S. He, Y. Jiang, and X. Ji, “Complementary attention for multi-agent reinforcement learning,” in International Conference on Machine Learning, pp. 30776–30793, PMLR, 2023.
\bibitem{CommNet}
S. Sukhbaatar, R. Fergus, et al., “Learning multiagent communication with backpropagation,” Advances in neural information processing systems, vol. 29, 2016.
\bibitem{REINFORCE}
R. J. Williams, “Simple statistical gradient-following algorithms for connectionist reinforcement learning,” Machine learning, vol. 8, no. 3, pp. 229–256, 1992.
\bibitem{IC3Net}
A. Singh, T. Jain, and S. Sukhbaatar, “Learning when to communicate at scale in multiagent cooperative and competitive tasks,” arXiv preprint arXiv:1812.09755, 2018.
\bibitem{bicnet}
P. Peng, Y. Wen, Y. Yang, Q. Yuan, Z. Tang, H. Long, and J. Wang, “Multiagent bidirectionally-coordinated nets: Emergence of human-level coordination in learning to play starcraft combat games,” arXiv preprint arXiv:1703.10069, 2017.
\bibitem{tarmac}
A. Das, T. Gervet, J. Romoff, D. Batra, D. Parikh, M. Rabbat, and J. Pineau, “Tarmac: Targeted multi-agent communication,” in International Conference on machine learning, pp. 1538–1546, PMLR, 2019.
\bibitem{dtmac}
C. Xu, H. Zhang, and Y. Zhang, “Multi-agent reinforcement learning with distributed targeted multi-agent communication,” in 2023 35th Chinese Control and Decision Conference (CCDC), pp. 2915–2920, IEEE, 2023.
\bibitem{maic}
L. Yuan, J. Wang, F. Zhang, C. Wang, Z. Zhang, Y. Yu, and C. Zhang, “Multi-agent incentive communication via decentralized teammate modeling,” in Proceedings of the AAAI conference on artificial intelligence, vol. 36, pp. 9466–9474, 2022.
\bibitem{magic}
Y. Niu, R. R. Paleja, and M. C. Gombolay, “Multi-agent graph-attention communication and teaming.,” in AAMAS, vol. 21, p. 20th, 2021.
\bibitem{imgs-mac}
S. Karten, M. Tucker, S. Kailas, and K. Sycara, “Towards true lossless sparse communication in multi-agent systems,” arXiv preprint arXiv:2212.00115, 2022
\bibitem{sukhbaatar}
S. Sukhbaatar, R. Fergus, et al., “Learning multiagent communication with backpropagation,” Advances in neural information processing systems, vol. 29, 2016.
\bibitem{standen}
M. Standen, M. Lucas, D. Bowman, T. J. Richer, J. Kim, and D. Marriott, “Cyborg: A gym for the development of autonomous cyber agents,” arXiv preprint arXiv:2108.09118, 2021.
\bibitem{pymarl2}
PyMARL 2, 2021, accessed: 2025-07-08. [Online]. Available: https://github.com/hijkzzz/pymarl2
\bibitem{wiebe}
J. Wiebe, R. A. Mallah, and L. Li, “Learning cyber defence tactics from scratch with multi-agent reinforcement learning,” arXiv preprint arXiv:2310.05939, 2023.
\bibitem{cymarl_commformer_github}
CyMARL-CommFormer, 2025, accessed: 2025-12-03. [Online]. Available: https://github.com/Poly-AIvsAI/CyMARL-CommFormer.

\end{thebibliography}

\end{document}